# A CRITICAL REASSESSMENT OF EVOLUTIONARY ALGORITHMS ON THE CRYPTANALYSIS OF THE SIMPLIFIED DATA ENCRYPTION STANDARD ALGORITHM

Fabien Teytaud[1] and Cyril Fonlupt[1]

[1]University Lille Nord de France, Calais, France
teytaud@lisic.univ-littoral.fr
fonlupt@lisic.univ-littoral.fr

## ABSTRACT

*In this paper we analyze the cryptanalysis of the simplified data encryption standard algorithm using meta-heuristics and in particular genetic algorithms. The classic fitness function when using such an algorithm is to compare n-gram statistics of a the decrypted message with those of the target message. We show that using such a function is irrelevant in case of Genetic Algorithm, simply because there is no correlation between the distance to the real key (the optimum) and the value of the fitness, in other words, there is no hidden gradient. In order to emphasize this assumption we experimentally show that a genetic algorithm perform worse than a random search on the cryptanalysis of the simplified data encryption standard algorithm.*

## 1. INTRODUCTION

Meta-heuristics are powerful tools for solving optimization problems. They have been applied to many combinatorial problems, and they are able to tackle problems for which the search space is too large for an exhaustive search. Cryptanalysis is the process of recovering a plain-text from a cipher. In order to deter attacks, the process relies on a large search space. For this reason, many meta-heuristics, and in particular Genetic Algorithm (GA) [1] [2], Memetic algorithm (MA), Simulated Annealing (SA) or Tabu search (TS) have been tried for cryptanalysis.

The Simplified Data Encryption Standard (SDES) is a simplified version of the well known Data Encryption Standard (DES) algorithm. The SDES has been designed for academic purposes and is used as a benchmark for cryptanalysis [3].

It is well known that cryptanalysis of the SDES scheme is an NP-hard problem and that meta-heuristics are well designed to solve combinatorial and difficult problems. By exploring a large set of solutions that improve over time, evolutionary algorithms have been successful for solving difficult and challenging problems. Even if the SDES is an academic and fairly easy problem that can be solved with an exhaustive search (as the key length is only 10 bits, there are no more than $2^{10} = 1024$ keys to try) it is used as a starting example for meta-heuristics and evolutionary cryptanalysis.

Currently, researchers in the cryptanalysis field look for regularities in the plain-text (if available), in the encrypted text, try to exploit vulnerabilities in the encryption process and stochastic search-based methods were not regarded as possible alternatives for cryptanalysis.

However as the brute force algorithms are inadequate for cryptanalysis for standard encryption schemes (DES requires a 56-bit length key for instance), meta-heuristics and evolutionary methods have drown a fair amount of attention from the cryptanalysis field.

As early as 1993, Spillman [4] was the first to introduce an evolutionary approach based on genetic algorithm for cryptanalysis to discover a simple substitution cipher. Mathhews [5] used genetic algorithms for transposition ciphers. In this work, GAs were used to seek the accurate permutation. In the same way, Jacobsen [6] proposed a hill-climbing approach to attack simple and polyalphabetical substitution ciphers.

More recently, several interesting studies were carried out for cryptanalysis of SDES via meta-heuristics and evolutionary approaches. Rao et al [7] were the first ones to study how several optimization heuristics (tabu search algorithms, simulated annealing and genetic algorithm) could match 9 to 10 bits of the target in about 15 to 20 minutes. No algorithms performed better than the others. This work was enhanced by Garg in several papers [8] [9] [10] [11] where she studied how to use memetic and genetic algorithms to break the SDES key. Other works regarding cryptography using evolutionary tools can be found in [12][13][14].

All the work described in the previous section provides a test-bed for evaluating the performance of memetic and evolutionary approaches and furthermore showed that unlike classic methods for breaking encryption schemes that require mathematical knowledge, these new approaches were attractive as they need little cryptological knowledge. However, we strongly believe that these approaches are biased and that on average they perform no better than random search and that the use of evolutionary schemes to break state-of-the-art encryption process require a better insight understanding of the encryption process.

The paper is organized as follows: Section 2 presents the SDES algorithm. Section 3 presents the random search while section 4 briefly describes the genetic algorithms. Experiments and discussions about the possible flaws of previous papers are explained in Section 5. Section 6 draws some conclusions and presents some hints to efficiently use memetic and evolutionary algorithms in cryptanalysis.

## 2. THE SIMPLIFIED DATA ENCRYPTION STANDARD

The SDES algorithm [3] is a simple encryption algorithm, it was devised for pedagogical purposes. It is a symmetric-key algorithm which means that the sender and the recipient share the same key.

It uses a 10-bit key and takes an 8-bit block of plain-text to produce an 8-bit block of cipher text. The decryption process works similarly, except the cipher is the input and the produced plain-text is the output.

The algorithm consists in five steps: an initial permutation (IP), a complex function $f_k$, another permutation function (SW), another application of the $f_k$ function and eventually a final permutation ($IP^{-1}$). This final permutation is the inverse of the initial permutation. These different steps are now detailed.

## 2.1 Key Generation

SDES is based on the use of a 10-bit key. From this key, two 8 bit sub-keys are produced respectively called $K_1$ and $K_2$. These two sub-keys are produced through different and easy binary operations: circular left shift and permutations.

## 2.2 Initial And Final Permutations

As previously said, the algorithm takes as input an 8-bit block. The first operation is an initial permutation called IP. The initial permutation is (1,5,2,0,3,7,4,6) which means that the 5th bit will be the 2nd bit after this initial permutation.

For instance, if we define the 8-bit block as $(m_0, m_1, m_2, m_3, m_4, m_5, m_6, m_7)$. Then, IP$(m_0, m_1, m_2, m_3, m_4, m_5, m_6, m_7)$ = $(m_1,m_5,m_2,m_0,m_3,m_7,m_4,m_6)$.

## 2.3 The $F_k$ Function

The $f_k$ function can be seen as the heart of the simplified data encryption standard SDES scheme, it is a complex function that involves a combination of permutation and non linear functions called Sboxes.

The $f_k$ function can be summed up as:
$$f_k(L, R) = (L \oplus F(R, SK), R)$$

Where F is a mapping from a 4-bit strings to a 4-bit strings, L is the leftmost 4 bits, R the rightmost 4 bits and SK the subkey. The mapping F is the trickiest part of the scheme as it involves non linear functions.

The first step is known as an expansion/permutation operation. It basically consists in mapping from a 4-bit string to an 8-bit strings.
For instance if the 4-bit input is $(m_0, m_1, m_2, m_3, m_4)$, the expansion/permutation operation consists in: $(m_4, m_1, m_2, m_3, m_2, m_3, m_4, m_1)$.

An exclusive or (xor) is performed on the output of the expansion/permutation operation with the first sub-key $K_1$:
$(m_4 \oplus k_{1,1}, m_1 \oplus k_{1,2}, m_2 \oplus k_{1,3}, m_3 \oplus k_{1,4}, m_2 \oplus k_{1,5}, m_3 \oplus k_{1,6}, m_4 \oplus k_{1,7}, m_1 \oplus k_{1,8})$ where $k_{1,i}$ indicates the ith bit of the sub-key.

This output is usually depicted for clarity reasons in a matrix form:

$$M = \begin{vmatrix} m_4 \oplus k_{1,1} & m_1 \oplus k_{1,2} & m_2 \oplus k_{1,3} & m_3 \oplus k_{1,4} \\ m_2 \oplus k_{1,5} & m_3 \oplus k_{1,6} & m_4 \oplus k_{1,7} & m_1 \oplus k_{1,8} \end{vmatrix}$$

At this step, the first row of this matrix M is used as input to the $S_0$ box to produce a 2-bit output while the second row is used to produce another 2-bit output from the $S_1$ box.

$$S_0 = \begin{bmatrix} 1 & 0 & 3 & 2 \\ 3 & 2 & 1 & 0 \\ 0 & 2 & 1 & 3 \\ 3 & 1 & 1 & 2 \end{bmatrix} \quad S_1 = \begin{bmatrix} 0 & 1 & 2 & 3 \\ 2 & 0 & 1 & 3 \\ 3 & 0 & 1 & 0 \\ 2 & 1 & 0 & 3 \end{bmatrix}$$

The process is quite easy. The 1st and 4th bits are turned into an integer to specify the row of the $S_0$ while the 2nd and 3rd bits are used to specify the column. For instance, if the first row of M is (1,0,1,0), the output of the $S_0$ box will be 2 (row 2 (10), column 1 (01)). This value is eventually turned into a binary number and delivers 2 bits. Additionally, the last row of M is used in the same way that $S_0$ as an input for $S_1$ to produce two more bits that are merged (the two bits of $S_0$ on the left, two bits of $S_1$ on the right).

These 4 bits undergo a last permutation called P4 ($m_2$, $m_4$, $m_3$, $m_1$) to provide the output the F function.

A final xor operation is performed on the output of the F function with the 4 leftmost bits of the input, the 4 rightmost bits being untouched.

As it can be seen, the output of the $f_k$ function only alters the leftmost 4 bits of the input. The purpose of the Switch function is to invert the input to iterate the $f_k$ function with the rightmost 4 bits.

## 2.4 The Switch Function

The switch function is a relatively simple function. It simply permutes the first four bits with the last four bits: SW($m_0$, $m_1$, $m_2$, $m_3$, $m_4$, $m_5$, $m_6$, $m_7$) = ($m_4$,$m_5$,$m_6$,$m_7$,$m_0$,$m_1$,$m_2$,$m_3$).

After this step, another iteration of the $f_k$ function is performed allowing to encrypt the 4 rightmost bits. However during this step, a slight modification is realized. Instead of xoring the output of the expansion/permutation operation with the sub-key $K_1$, the sub-key $K_2$ is used instead.
A summary of the SDES algorithm is presented in Figure 1.

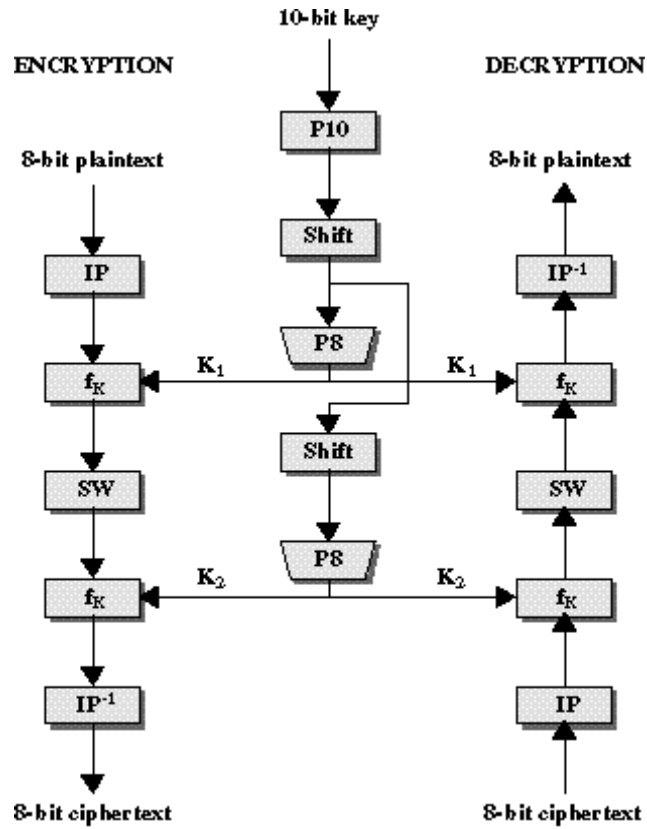

**Figure 1**. The SDES algorithm[1].

## 3. RANDOM SEARCH

Random Search (RS) belongs to the family of numerical optimization. It does not require the gradient of the function to be optimized and hence, can be performed on functions which are not continuous or differentiable. Such optimization can be applied to black-box optimization. The principle of this method is relatively simple : at each iteration a random solution from the search space is generated. If this solution is better than the current best solution, this solution replaces the current best solution. At the end, the best solution found by the algorithm is returned.

Needless to say, this method is a basic method which is studied here only to enhance the problematic of this paper. This method can be seen as a lower bound of the quality of a search algorithm.

## 4. GENETIC ALGORITHMS

Evolutionary algorithms belong to the family of stochastic optimization algorithms. These methods are bio-inspired techniques that crudely mimic reproduction, mutations, crossovers and selection. They are modeled according to Darwin's evolution theory. In these algorithms, individuals represent candidate solutions of the optimization problem. These algorithms do not require the gradient of the problem to be optimized. Genetic Algorithms (GA) [1] [2] belong to the family of evolutionary algorithms. They are mainly used with a discrete search space, meaning that they are used for combinatorial optimization problems. A population of candidate solutions evolves, and generations after generations, individuals are biased towards the best ones (according to the fitness function). Crossovers and mutations are random operators used for exploring the search space. Algorithm 1 illustrates this method.

---

1  Figure from *http://homepage.smc.edu/morgan_david/vpn/des.htm*

This algorithm is used in Section 5 as it is praised in [9] as a good scheme to hack the key of SDES.

```
1.  Generate the initial parent Population
2.  Evaluate all individuals of the parent Population
3.  While a stopping condition is not reached do
4.      for all individuals i in population do
5.          parent1 = parentalSelection(parentPopulation)
6.          if crossover probability is satisfied then
7.              parent2 = parentalSelection(parentPopulation)
8.              offspringPopulation[i] = crossover(parent1, parent2)
9.          else
10.             offspringPopulation[i] = parent1
11.         endif
12.         if mutation probability is satisfied then
13.             offspringPopulation[i] = mutate(offspringPopulation[i])
14.         endif
15.     endfor
16.     Evaluate all individual of the offspringPopulation
17.     parentPopulation=
            survivalSelection(offspringPopulation, parentPopulation)
18. endwhile
```

**Algorithm 1.** A generic Genetic Algorithm.

## 5. EXPERIMENTS

### 5.1 Checking The Encryption Process Performance

Using bio-inspired algorithms for the cryptanalysis of the simplified DES is based on a very strong assumption: a fitness function can guide the search towards the perfect encryption key.

The technique that is currently used for any language (in this paper, English is used without loss of generality) is to compare n-gram statistics of the decrypted message with those of the target message.

To evaluate the suitability of a proposed key K, the encrypted text is decrypted using this test key K, and the statistics of the decrypted text are compared to the statistics of the target language. Usually, only unigrams and bigrams are used since trigrams require more computing power and time. This additional cost is too important in comparison with its corresponding information.

This fitness function is defined as follows:

$$F_k = \alpha \sum_i |E_i - D_i| + \beta \sum_i |E_{(i,j)} - D_{(i,j)}| \quad (1)$$

with α and β being the weights given to unigram and digram,

where $E_i$ is the percentage of the ith letter in the English language and $E(i,j)$ is the frequency of a letter i followed by a letter j in English (for instance the frequency of "th" is 1.52 while the frequency of "ld" is 0.02). $D_i$ is the frequency of a letter i in the decrypted text and $D(i,j)$ is the frequency of the digram ij in the decrypted message.

This fitness function is 0 if the correct key K is used in the decryption process and the greater the value of $F_k$, the worst the solution.

It is quite obvious that this fitness function can be used in a minimization process and when this function tends towards 0 we can assume that the text is close to the plain-text. It is a good practice to evaluate this fitness functions over several texts and the longer the text is, the more accurate the statistics of unigram and digram are. In the case of genetic or memetic algorithms, this fitness function is used to evaluate candidate key and to guide the evolutionary process towards the good solutions (keys).

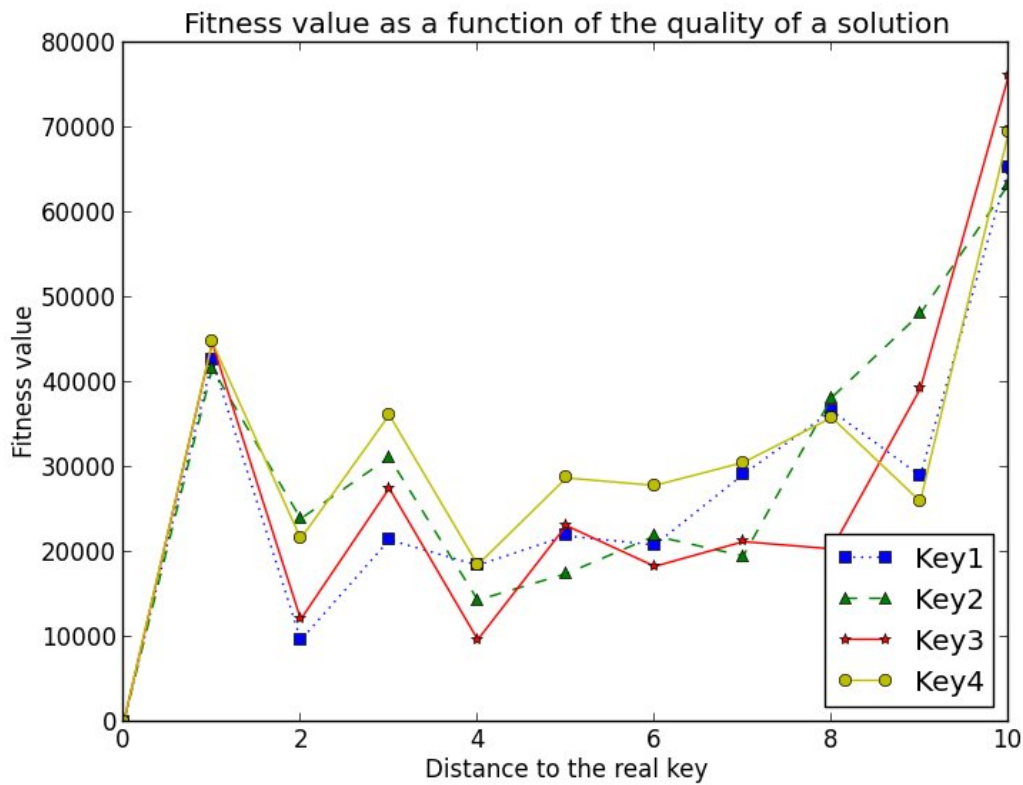

*Figure 2. Relevance of the fitness function*

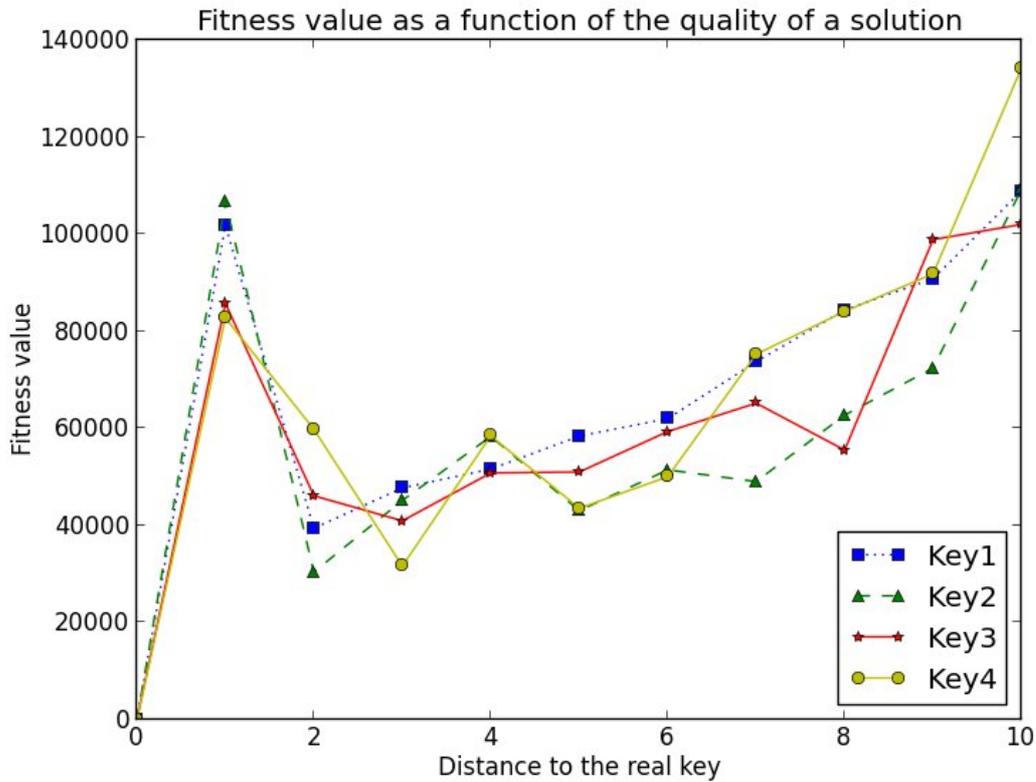

***Figure 3.*** *Relevance of the fitness function using only unigrams*

For methods like genetic algorithms, memetic algorithms or hill-climbing strategies, a strong assumption is made: there exists an implicit gradient to better solutions to the best solution. In other words this means that the more bits a solution shares with the real key (*i.e.* the target key), the greater the fitness value, thus guiding the research to share more bits with the real key. In [9] [13], Garg shows that using memetic algorithm, several bits of the key are correctly recovered after several minutes for cipher text ranging from 100 to 1000 characters. However, we think that there is some serious flaw in the reasoning: the fact that this fitness function is fruitful to the search landscape.

In order to show that using the fitness function $F_k$ to guide the evolutionary process is more or less like trying to find a needle in a haystack, we devise the following experiment: starting with a given key of 10 bits, we perturb one of the 10 bits of this original key to get a new key and we evaluate $F_k$ over this new key. This experiment is iterated 9 times, one for each possible key with a distance 1 from the original key, then we go on with all the possible keys with a distance of 2 from the original key and so on until we evaluate the fitness function for all possible keys with a distance of 9 (note that for all cases, the number of possible keys with a distance n is equal to the number of keys with a distance 10-n).

Figures 2 and 3 represent the fitness values of the perturbation of 8 random keys (4 per figure). Due to the lack of space, only 8 different keys are displayed but are fully representative of the

1024 different keys. From these figures the first point to note is that there is absolutely no correlation between any two perturbed keys whatever its distance from the original key is! Furthermore, it seems that solutions at one bit distance from the target key are attributed a misleading fitness value [15]

When comparing these figures with results from [9][10][13], we can understand why the memetic approach seems to remain stuck to no more than 9 bits matched (out of 10), the solutions at a distance of one bit of the local optimum act as a deterrent and hopping from these solutions to converge to the global optimum is rather intractable.

**5.2 Comparison Between A Random Search And A Genetic Algorithm.**

In the previous section, the fitness landscape of the SDES is examined and we show that is is quite deceptive. In this section, we study the number of bits matching with the target key for two methods: a GA-based approach and the classic random search scheme. To assess our approach and to show that using a GA with the fitness function presented is Section 5.1 and used in [11] for cryptanalysis is not an efficient tool for cryptanalysis, we carry out the following experiments.

As in [9], a simple GA is programmed whose fitness function is equation 1 and is compared to a random search method. The purpose of these experiments is to show whether or not a GA performs better or not than a random search for the cryptanalysis of SDES.

The GA is tuned as in [9], the population size was set to 10, the probability for crossover is set to 0,95 while the probability for mutation is 0.05. 10 generations are performed, so that the total number of evaluations is the same than the experiment using random search. Table 2 sums up the results for the GA approach for texts of various lengths (from 100 to +100k characters) while table 1 shows the results for the RS scheme. There is no specific parameters for the RS scheme: a solution is simply randomly generated. For each text size, 100 runs of the GA and RS are performed, the standard deviation and the average fitness of the number of matched bits are computed. The best of the 100 runs is recorded.

According to these results it is quite obvious that the GA does not outperform the RS. Moreover is seems that the GA heuristic seems to add some kind of deleterious mechanism to the search.

From all these results it seems clear that we can conclude that using a GA with this fitness function is irrelevant and that a basic random search, usually considered as a lower bound, is more efficient.

We can also say that this kind of problems is not suited to meta-heuristics like GA as the SDES like all cryptographic schemes is essentially devised as a misleading problem with no gradient. If some kind of gradient was available to the problem, SDES, like DES, would be much more easily hackable.

*Table 1. Results of the random search.*

| Text size | Fitness Mean | Standard deviation | Best fitness found |
|---|---|---|---|
| 100 | 5,4 | 2,27 | 10 |
| 200 | 6,4 | 1,57 | 10 |
| 400 | 6,9 | 1,52 | 10 |
| 800 | 6,7 | 1,49 | 10 |
| 1600 | 7,4 | 2,01 | 10 |
| 3200 | 6,3 | 1,34 | 8 |
| 6400 | 6,9 | 1,97 | 10 |
| 12800 | 6,7 | 2 | 10 |
| 25600 | 5,8 | 2,74 | 10 |
| 51200 | 6,4 | 2,72 | 10 |
| 102400 | 5,2 | 1,48 | 7 |

*Table 2. Results of the Genetic Algorithm.*

| Text size | Fitness Mean | Standard deviation | Best fitness found |
|---|---|---|---|
| 100 | 5,4 | 2,27 | 10 |
| 200 | 5,6 | 1,51 | 8 |
| 400 | 5,7 | 1,57 | 7 |
| 800 | 5,8 | 2,1 | 10 |
| 1600 | 4,9 | 1,66 | 8 |
| 3200 | 6,1 | 1,91 | 10 |
| 6400 | 6,4 | 1,90 | 9 |
| 12800 | 5,4 | 1,90 | 8 |
| 25600 | 6,3 | 2,16 | 10 |
| 51200 | 5,8 | 1,40 | 8 |
| 102400 | 5 | 1,89 | 8 |

To conclude, using an evolutionary approach for the cryptanalysis of SDES is not an adequate response. Moreover, it seems that it in some cases it performs worse than the random search.

# 6. CONCLUSIONS

In this paper we emphasize the importance of having a relevant fitness function when meta-heuristic methods, and in particular when genetic algorithms are used to the cryptanalysis of the SDES . We first study the landscape of the fitness function in order to show that there is no correlation between the fitness function and the distance to the correct key. This is an important fact because this means that having a good score (according to the fitness) does not mean that we are close to the optimum. Then we compare the efficiency of the genetic algorithm with a random search. From these experiments, we can conclude that the genetic algorithm does not converge correctly to the optimum. The SDES algorithm uses a 10-bit key, meaning the search space equals to 1024. This allows us to perform a random search. In fact the results is quite interesting: in average the random search is better than the genetic algorithm. The number of correct bits for all text sizes is 6.37 for the random search against 5.67 for the genetic algorithm. Moreover 8 times (over 10) the optimum (*i.e.* the correct key) is found with the random search against 4 times over 10 for the genetic algorithm. The main explication comes from the bad fitness function used. Figures 2 and 3 show in particular that being at a distance of 1 to the correct key corresponds generally to a fitness trap. There is no doubt that the genetic algorithm converges to a local optimum.

We strongly believe that using evolutionary computing techniques might be of help for cryptanalysis but it should be used not as a silver bullet to magically retrieve the key but more as the tool for cryptanalysis. The most complex part of SDES consists of a combination of the non-linear functions (called Sbox); these Sboxes might be built using evolutionary computation approaches and may also be hacked with this approach.

# REFERENCES
1: John Holland, Adaptation in natural and artificial systems: An introductory analysis with applications to biology, control, and artificial intelligence, 1975
2: David Goldberg, John Holland, , Genetic algorithms and machine learning, 1988
3: Edward F.Schaefer, A simplified Data Encryption Standard Algorithm, 1996
4: Richard Spillman, Cryptanalysis of Knapsack Cipher using Genetic Algorithms, 1993
5: Robert Mathews, The Use of Genetic Algorithms in Cryptanalysis, 1993
6: Thomas Jakobsen, A Fast Method for Cryptanalysis of Substituion Ciphers,
7: N. Nalini, G. Raghavendra Rao, Cryptanalysis of Simplified Data Encryption Standard via Optimisation Heuristics, 2006
8: Poonam Garg, Genetic Algorithm Attack on SImplified Data Encryption Standard Algorithm, 2006
9: Poonam Garg, A Comparison between Memetic algorithm and Genetic Algorithm for the cryptanalysis of Simplified Data Encryption Standard algorithm, 2009
10: Poonam Garg, Cryptanalysis of SDES via Evolutionary Computation Techniques, 2009

11: Poonam Garg, Evolutionary Computation Algorithms for Cryptanalysis: A Study, 2010
12: Mishra, Swati,Siddharth Bali, Public key cryptography using genetic algorithm., 2013
13: Sharma, Lavkush, Bhupendra Kumar Pathak,R. G. Sharma, Breaking of Simplified Data Encryption Standard Using Genetic Algorithm., 2012
14: Ratan Ram, Applications of Genetic Algorithms in Cryptology., 2014
15: David Goldberg, Bradley Korb, Kalyanmoy Deb, Messy genetic algorithms: motivation, analysis, and first results, 1989